# Use of negative capacitance to provide a sub-threshold slope lower than 60 mV/decade

Sayeef Salahuddin and Supriyo Datta

*School of Electrical and Computer Engineering and NSF Center for Computational Nanotechnology (NCN), Purdue University, West Lafayette, IN-47907*

*Abstract*

It is well known that conventional Field Effect Transistors (FET's) require a change in the channel potential of at least 60 mV at 300K to effect a change in the current by a factor of ten, and this minimum subthreshold slope S puts a *lower limit of fundamental nature* on the operating voltage and hence the power dissipation in standard FET based switches. Here we show that by replacing the standard insulator with a ferroelectric insulator of the right thickness it should be possible to implement a ***step-up voltage transformer*** that will amplify the gate voltage thus leading to values of S lower than 60 mV/decade and enabling low voltage/low power operation. The voltage transformer action can be understood intuitively as the result of an effective negative capacitance provided by the ferroelectric capacitor which arises from an internal positive feedback that in principle could be obtained from other microscopic mechanisms as well. Unlike other proposals to reduce S this involves no change in the basic physics of the FET and thus does not affect its current drive or impose other restrictions.

*Introduction:* It is generally accepted that the ongoing scaling of field effect transistors (FET's, see Fig.1) will be eventually limited by the inability to remove the heat generated in the switching process [1,2,3,4], making it very important to find ways to reduce the power dissipated per switching event. It is also clear that the power dissipation would be lowered significantly, if FET's could be operated at lower voltages. A key factor limiting the operating voltage is the subthreshold swing S, which is the inverse of the change of current that can be obtained for a unit change in gate voltage, $\mathbf{V}_g$ :

$$\mathbf{S} \equiv \frac{\partial \mathbf{V}_g}{\partial (\log_{10} \mathbf{I})} = \underbrace{\frac{\partial \mathbf{V}_g}{\partial \psi_s}}_{\equiv m} \frac{\partial \psi_s}{\partial (\log_{10} \mathbf{I})} \qquad (1)$$

Standard FET analysis shows that the second term $\partial \psi_s / \partial(\log_{10} \mathbf{I})$ relating the change in the current to the change in the surface potential in the channel cannot be any lower than 60 mV/decade at room temperature. Since $\mathbf{V}_g$ and $\psi_s$ are related by a capacitive voltage divider as shown in Fig.1, it is apparent that the first term $\partial \mathbf{V}_g / \partial \psi_s$ (often called the body factor 'm') given by

$$\frac{\partial \mathbf{V}_g}{\partial \psi_s} = 1 + \frac{C_s}{C_{ins}} \qquad (2)$$

must exceed one, thus putting a lower limit of 60 mV/decade (corresponding to m = 1) on the subthreshold slope S. Even high-$\kappa$ insulators [5] with phenomenally large values of $C_{ins}$ can only reduce the body factor 'm' so as to approach one, but cannot make it any smaller.

The objective of this paper is to show that if we replace the conventional insulator with a ferroelectric insulator, having a P-E (polarization versus electric field) characteristic of the type shown in Fig. 2, it is possible to obtain $\partial V_g / \partial \psi_s < 1$ and hence a value of S lower than 60 mV/decade. This can be understood from Eq.(2) by noting that the ferroelectric capacitor is effectively a **negative** one, since the slope of P versus E (which is a scaled version of Q versus V) around the origin is **negative.** Ordinarily this negative slope segment is unstable and not directly observed in experiments which exhibit hysteretic jumps in the polarization. But we argue below that if the ferroelectric capacitor is placed in series with a normal capacitor, the negative capacitance segment can be effectively stabilized making it possible for the channel potential $\psi_s$ on an internal node to change more than the voltage $V_g$ applied externally, thus providing a voltage amplifier or more correctly, a ***step-up voltage transformer.*** Although we use ferroelectrics to illustrate the concept, we show below that negative capacitance generally arises from a positive feedback that could in principle be obtained through other microscopic mechanisms as well.

Note that our proposal to reduce the first term in Eq. (1) (that is, the body factor, '*m*') is very different from other proposals to reduce S which usually assume that *m* cannot be reduced below one and instead seek to improve the second term ($\partial \psi_s / (\log_{10} I)$) by using a different transistor operating principle such as band-to-band tunneling [6] or impact ionization [7] leading to different current and voltage requirements that may offset the advantage gained in S. By contrast our proposal requires no change in the basic transistor operation. It simply inserts a step-up voltage transformer in the gate circuit that could help alleviate the device-interconnect "voltage mismatch" emphasized by Yablonovitch [8]. It should also be noted that our approach is very different from the well studied field of Ferroelectric RAM where the focus is on the memory devices [see for example Ref. 9 and references therein] and Ferroelectic FET where the motivation is to use ferroelectric oxide as a high-$\kappa$ gate material [see for example Ref. 10 and references therein].

*Negative capacitance from positive feedback:* The negative capacitance in ferroelectrics can be understood in terms of a positive feedback [11] mechanism as follows. Suppose we have a (positive) capacitor $C_0$ (per unit area) which sees a terminal voltage equal to the applied voltage V plus a feedback voltage $\alpha_f Q$ proportional to the charge on the capacitor Q (per unit area), such that

$$Q = C_0 (V + \alpha_f Q)$$

This yields $Q = C_{ins} V$, where $$C_{ins} = \frac{C_0}{1 - \alpha_f C_0} \qquad (3)$$

Clearly with $\alpha_f C_0 > 1$, we have a negative capacitance which would ordinarily lead to an instability so that the charge would increase till limited by the non-linear terms that we have neglected so far. If we stabilize the negative capacitor by putting an ordinary capacitor $C_s$ in series such that the overall capacitance $[C_s^{-1} + C_{ins}^{-1}]^{-1}$ is positive, then it follows from Eqs.(2) and (3) that the body factor

$$\frac{\partial V_g}{\partial \psi_s} = 1 - \frac{C_s}{C_0}(\alpha_f C_0 - 1) \qquad (4)$$

can, in principle, be made arbitrarily small showing that the negative capacitance can be used as a voltage transformer that steps up the applied potential of $V_g$ into a channel potential of $\psi_s$. As long as $C_{eq}^{-1} = C_s^{-1} + C_{ins}^{-1}$ remains positive, the composite ferroelectric-semiconductor system behaves like a normal positive capacitor.

More generally, we could replace the linear capacitor $Q = C_0(V + \alpha_f Q)$ with a general nonlinear capacitance function $Q = F(V + \alpha_f Q)$, such that

$$\mathbf{V} = \mathbf{F}^{-1}(\mathbf{Q}) - \alpha_f \mathbf{Q}$$

Noting that the function $F^{-1}$ is generally odd, we could expand it approximately up to the fifth power (the first term being the inverse linear capacitance $1/C_0$) to write

$$V \approx \alpha_0 Q + \beta_0 Q^3 + \gamma_0 Q^5 + \rho_0 \frac{dQ}{dt} \qquad (5)$$

where we have also added a possible resistive drop proportional to the current dQ/dt. Note that the parameter $\alpha_0$ equals $(1/C_0) - \alpha_f$ and will be negative if we are in the negative capacitance regime where $\alpha_f C_0 > 1$.

*Ferroelectric capacitor:* Above we have tried to illustrate the generic role of positive feedback in giving rise to negative capacitance, but it should be noted that Eq.(5) does not just represent a toy model for ferroelectrics. It can just as well be obtained starting from the state-of-the-art approach for modeling the dynamics of ferroelectric capacitors based on the Landau-Khalatnikov (LK) equation [12,13,14]

$$\rho \frac{d\vec{P}}{dt} + \vec{\nabla}_P U = 0 \tag{6}$$

$$\text{where} \quad U = \alpha P^2 + \beta P^4 + \gamma P^6 - \vec{E}_{ext} \cdot \vec{P} \tag{7}$$

is the Gibb's free energy given by the sum of the anisotropy energy and the energy due to the external field $\vec{E}_{ext}$, and $\vec{P}$ is the polarization charge per unit area. From Eqs.(6) and (7)

$$E_{ext} = 2\alpha P + 4\beta P^3 + 6\gamma P^5 + \rho \frac{dP}{dt} \tag{8}$$

where we have dropped the vector signs assuming the spatial variations to be one-dimensional. It is easy to see that if we set Q = P and V = $E_{ext} t_{ins}$ in Eq.(8) where $t_{ins}$ is the insulator thickness, we obtain Eq.(5) with $\alpha_0 = 2\alpha t_{ins}$, $\beta_0 = 4\beta t_{ins}$, $\gamma_0 = 6\gamma t_{ins}$ and $\rho_0 = \rho t_{ins}$.

The plot of the steady-state polarization P versus $E_{ext}$ shown earlier (Fig.2) was obtained from Eq.(8) setting dP/dt =0 and using parameters typical of $BaTiO_3$: $\alpha = -1 e7 \, m/F$, $\beta = -8.9 \, e9 \, m^5/F/coul^2$, $\gamma = +4.5 \, e11 \, m^9/F/coul^4$ [15]. The negative slope for small P arises from the negative values of $\alpha$ and $\beta$ which get suppressed by the positive $\gamma$ at larger P values. Note that an ordinary linear dielectric with a dielectric constant $\varepsilon$ would have a positive $\alpha = 1/2\varepsilon$, with $\beta=0$ and $\gamma=0$.

*Ferroelectric capacitor in series with an ordinary capacitor, $C_s$:* Let us now assume that the FET shown in Fig.1 has a ferroelectric insulator rather than an ordinary insulator at the gate. The gate circuit can be assumed to be a series combination of the ferroelectric capacitor and another capacitor that we will call $C_s$ representing semiconductor, channel-to–source and channel-to-drain capacitances. For this series combination, a voltage $\psi_s$ appears across $C_s$, while the rest $\mathbf{V}_g - \psi_s$ appears across the ferroelectric such that both have the same charge Q and we can write

$$\psi_s = Q/C_s \tag{9a}$$

$$\mathbf{V}_g - \psi_s \approx \alpha_0 Q + \beta_0 Q^3 + \gamma_0 Q^5 + \rho_0 \frac{dQ}{dt} \tag{9b}$$

making use of Eq.(5). Combining these expressions we can relate the applied voltage $\mathbf{V}_g$ to the potential $\psi_s$ that appears inside the channel:

$$\tau \frac{d\psi_s}{dt} + (1+a_1)\psi_s + a_2\psi_s^3 + a_3\psi_s^5 = \mathbf{V}_g \qquad (10a)$$

where $\tau = \rho C_s t_{ins}$, $a_1 = 2\alpha C_s t_{ins}$, $a_2 = 4\beta C_s^3 t_{ins}$ and $a_3 = 6\gamma C_s^5 t_{ins}$ (10b)

***Steady-state response:*** The steady-state $\psi_s$ versus $\mathbf{V}_g$ is obtained from Eq.(10a) by setting $d\psi_s/dt = 0$:

$$(1+a_1)\psi_s + a_2\psi_s^3 + a_3\psi_s^5 = \mathbf{V}_g \qquad (11)$$

Obviously the nature of the $\psi_s$ vs. $\mathbf{V}_g$ characteristics will depend on the three coefficients $a_1$, $a_2$ and $a_3$ which are scaled versions of the material parameters $\alpha$, $\beta$ and $\gamma$ (see Eq.(10)). If the parameter values are such that, the non-linear terms $a_2$ and $a_3$ are very small compared to $a_1$, then the relation between $\psi_s$ and $\mathbf{V}_g$ is essentially linear:

$$\partial \mathbf{V}_g / \partial \psi_s \approx 1 + a_1$$

which is exactly the same as our earlier Eq.(4), as we might expect since we are neglecting the non-linear terms. But the point to note is that even if $\boldsymbol{a}_2$ and $\boldsymbol{a}_3$ are non-negligible, $(1+\boldsymbol{a}_1)$ represents the slope close to the origin $\psi_s=0, \mathbf{V_g}=\mathbf{0}$ and should be positive if we wish the origin to represent a stable operating point. Using Eq.(10b) for $a_1$ it is easy to see that the condition $(1+a_1) > 0$ requires the thickness of the ferroelectric insulator to be less than a critical thickness defined as

$$t_{ins} \leq \frac{1}{2|\alpha|C_s} \equiv t_c \tag{12}$$

This condition ensures that the ferroelectric capacitor is large enough (or the series capacitor $C_s$ is small enough) that the combination forms a stable positive capacitor, and there is no hysteresis at the origin unlike Fig.2 which corresponds to a very thick ferroelectric. All results we present below assume $t_{ins} < t_c$, since a hysteresis at the origin seems undesirable for device applications.

Fig.3 shows $\psi_s$ vs. $\mathbf{V}_g$ for different values of $t_{ins}$, obtained directly from Eq.(11) using appropriate values of $\alpha$, $\beta$ and $\gamma$ for $BaTiO_3$, in series with a linear capacitance $C_S = 100\, fF/\mu m^2$ [16]. The plots are obtained starting from an initial state with the ferroelectric capacitor fully negatively polarized, sweeping the voltage towards the positive maximum and then sweeping it back again to the negative maximum. Note that the insulator thicknesses we use are less than the critical thickness which equals 500 nm for this choice of parameters.

As the insulator thickness is increased, the plots become progressively steeper but at the same time, they start to show more and more hysteresis away from the origin (but not at the origin) due to the nonlinear terms $a_2$ and $a_3$. Fig. 4 shows the corresponding plots for $\Delta\psi_s/\Delta V_g$ as a function of insulator thicknesses for a change in $V_g$ from 0 to 0.2V. Note that in this voltage range the actual $\partial\psi_s/\partial V_g$ for specific voltages is many times more than the $\Delta\psi_s/\Delta V_g$ that we have shown. Fig.4 shows that with $t_{ins}$ = 250 nm, a voltage gain > 4 is possible resulting in a channel potential $\psi_s > 0.8$V for an applied gate voltage $V_g$ of 0.2 V. Fig.3 shows that with this insulator thickness there is a hysteresis in $\psi_s$ vs. $V_g$ (away from the origin), but it is small. In general a trade off may be needed between steepness and hysteresis, unless materials can be engineered to minimize the nonlinear term $\beta$ relative to the linear term $\alpha$ so as to avoid hysteresis even at large steepness (see appendix).

It will be noted from Eq. (10 b) that $C_s$ were smaller, the non linear terms $a_2$ and $a_3$ would become smaller in comparison to the linear term $a_1$ making the hysteresis negligible in our voltage range of interest. The value for $C_s$ that we have used in this paper is more appropriate for ultra small FETs [16], where the short channel effects are significant. For a present day commercial device, the value of $C_s$ should be roughly a factor of 10 lower than that used in this paper and as such should show steep $\psi_s$ vs. $V_g$ with negligible hysteresis. However, since the critical thickness is inversely proportional to $C_s$ (see Eq. 12), the thickness of the ferroelectric insulator will have to be roughly a factor of 10 larger than those shown in Fig. 3.

***Concluding remarks:*** In summary, we have shown that any microscopic mechanism that can provide the positive feedback needed to result in a negative capacitance, can be used to implement a step-up voltage transformer that would allow low voltage/low power operation of conventional FET's. For a ferroelectric capacitor, it is the dipole interaction that provides the positive feedback. Other mechanisms such as avalanche breakdown, polaronic effect etc. within the oxide can possibly provide the positive feedback needed for negative capacitance as well. The parameters $\alpha_0, \beta_0, \gamma_0$ and $\rho_0$ in Eq.(5) will be determined by the specific mechanism involved.

Negative capacitance regions (as in Fig.2) are ordinarily unstable and not observed in experiments. But placing such a capacitor in series with a positive capacitor stabilizes it by making the effective $\alpha$ of the composite capacitor positive, provided its thickness is less than the critical thickness defined in Eq.(12). Since $\alpha$ is proportional to (T-$T_c$), $T_c$ being the Curie temperature, the series combination acts like a ferroelectric at a temperature above its $T_c$. Indeed the plots in Fig.3 look very similar to the experimental results reported for P versus E in the classic work of Merz [17, see Figure 1] for ferroelectrics at temperatures around $T_c$.

The LK equation used in this paper assumes a single domain ferroelectric insulator. However, the effect is not limited to single domain ferroelectric materials. Rather, any ferroelectric material showing a 'run-away' effect in switching such that once a domain is nucleated, the others follow suit, should show similar negative capacitance effect. It is the 'run-away' process or positive feedback that is important for the negative capacitance to manifest itself. In fact, the values of $\alpha$, $\beta$ and $\gamma$ that we have used in this paper were extracted by Merz [17] by fitting Eq. (7) to experimental data where the switching was clearly mediated by domain nucleation. In general, materials with $\alpha$ large enough to make $\beta$ negligible are preferred as their behavior will approach the ideal negative capacitance described by Eq. (4).

It is evident from Eqs.(10 and (12) that the time scale for switching is theoretically given by $\tau = \rho C_s t_{ins} \leq \rho/2|\alpha|$. Experimentally intrinsic switching times of the order of 70~90 picoseconds have been reported [18], but it remains to be seen how much the speed can be improved by appropriate materials engineering.

In a recent paper [19], we used the Landau-Lifshitz-Gilbert (LLG) equation to show that it should be possible to switch a collection of N interacting spins that act in concert with considerably less energy than that needed to switch N non-interacting spins. A similar physics is possibly at work here with the dipoles in a ferroelectric acting in unison leading to a reduction in the switching energy.

To conclude, we have shown that it is possible to reduce the sub threshold swing of a FET below the theoretical limit of 60 mV/decade by using a ferroelectric insulator in the gate. The ferroelectric provides a negative capacitance that should make it possible to implement a step-up voltage transformer reducing the subthreshold swing below 60 mV/decade limit, enabling low voltage/low power operation. This negative capacitance is a result of the positive feedback among the electric dipoles in the ferroelectric and in principle can be obtained from other microscopic principles as well.

*Acknowledgements:* It is a pleasure to acknowledge helpful discussions with Mark Lundstrom, Ashraf Alam and Tony Arrott. This work was supported by the Nanoelectronics Research Initiative (NRI).

*Appendix:* In this section, we shall look at two things namely an alternative description of the Fig.3 and Fig.4 and the effects of the ferroelectric parameters such as $\alpha$ and $\beta$ on the overall device performance. In particular, we show that a load line analysis of the series combination of the ferroelectric capacitor and linear capacitor yields the same results as described by Eq. (9) and (10) and shown in Fig. 3 and 4. Furthermore, by looking at the parameter variation of the ferroelectric capacitor, we show that the most desirable behavior is obtained when $\alpha$ is large enough so that the influence of the non linear terms $\beta$ and $\gamma$ can be ignored. Hence material optimization should be focused on increasing $\alpha$.

**Load Line Description of the problem:**

It is evident from Eq. (9) that linear capacitor acts as a load line of negative slope on the S-shaped curve of Fig 2, which goes through the points ($Vg/t_{ins}$, 0) and (0, $CVg$). This is shown in Fig. S1. If the slope of the line is very high, i.e. the capacitance is small, then the load line cuts through the negative slope region of the S-curve (see the left panel), as a result $\psi_s$ goes through zero. On the other hand if the capacitance is larger (see the right panel) the load line can not cut the negative slope region and hence the switching occurs abruptly. This explains the particular shapes obtained in Fig. 3 and 4.

**Effects of parameter variations:**

In Fig. 3 and 4 of the main article we have shown the behavior of $\psi_s$ as a function of gate voltage $\mathbf{V_g}$ for a set of ferroelectric parameters appropriate for bulk BaTiO$_3$ [15]. However, for ferroelectric thin films, these values may change. In addition, for device optimization we should have a good understanding of how these parameters will affect the overall behavior. In the following we look at the effects of variation of $\alpha$ and $\beta$. The other parameter $\gamma$, being positive, essentially adds the non linearity that restricts polarization to increase beyond bounds and will not be discussed here.

**Effect of increasing $\alpha$:**

Increasing $\alpha$ circumvents the nonlinear term and the ferroelectric oxide approaches closer to an ideal negative capacitance. This can be seen from the middle panel of Fig. S2. We see that a huge gain in $\Delta\Psi_s/\Delta\mathbf{V_g}$ is possible with almost negligible hysteresis. Hence an increased $\alpha$ will help device performance. We note that the switching field for the isolated ferroelectric increases. Also, due to the fact that an increasing $\alpha$ effectively decreases the capacitance of the ferroelectric oxide, we need a smaller thickness of the insulator compared to Fig. 3 and 4 for hysteresis to appear in $\psi_s$ vs. $V_g$.

**Effect of increasing** $\beta$ **:**

Figure S3 shows the effect of decreasing $\beta$ with $\alpha$ and $\gamma$ fixed to their BaTiO$_3$ values. We see that the switching field is decreased. Also the gain in $\Delta\Psi_s/\Delta V_g$ is not that much even when the hsyteresis is such that the $\psi_s$ vs. $V_g$ plot does not go through zero.

From the above analysis, we see that the most desirable behavior, i.e., a large $\partial\psi_s/\partial V_g$ with small hysteresis occurs when $\alpha$ is large. Hence, i*n general, we want as big an* $\alpha$ *as possible.*

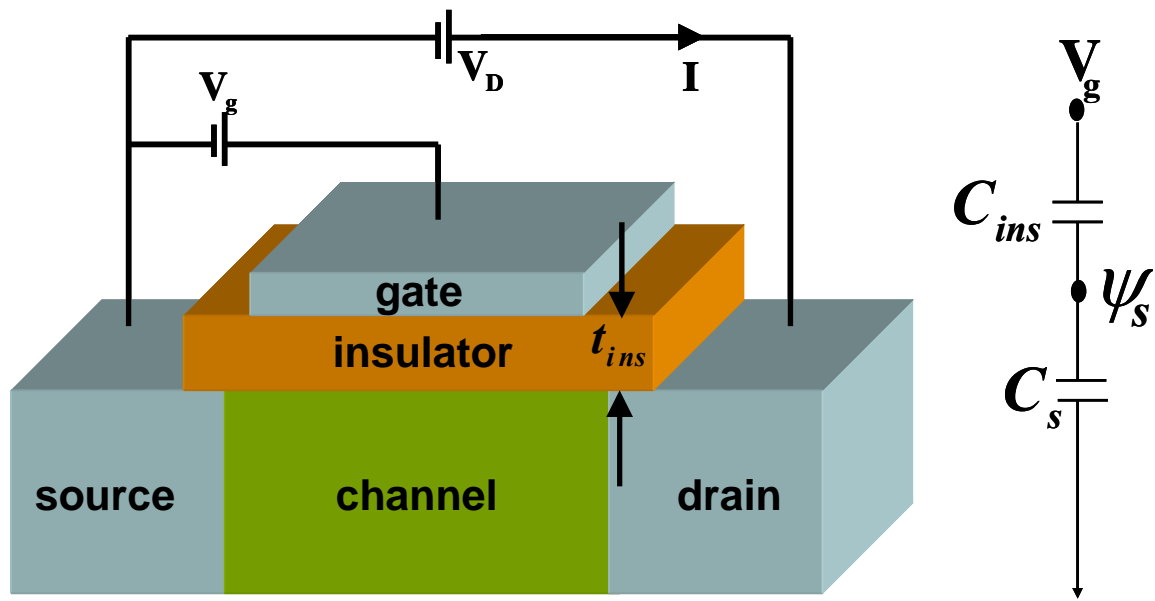

**Fig. 1:** A standard Field Effect Transistor (FET) structure where the current I in the drain circuit is controlled by the gate voltage $V_g$. The right panel shows an equivalent circuit for the division of the gate voltage between the insulator capacitance and the semiconductor capacitance (that comprises of the delpletion, channel to source and channel to drain capacitances).

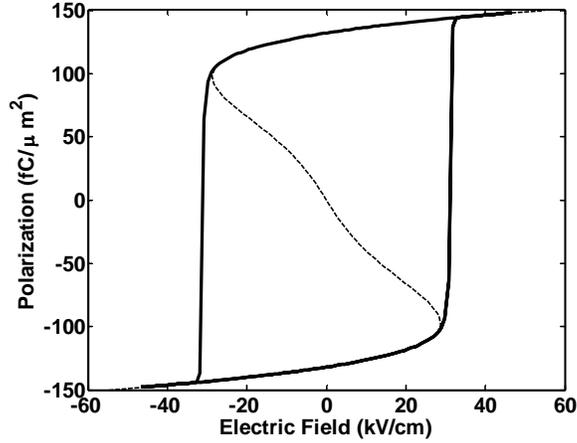

**Fig. 2:** Polarization versus electric field for a typical ferroelectric material calculated from the LK equation (see Eq.(1)) using parameters appropriate for $BaTiO_3$: $\alpha = -1\,e7\,m/F$, $\beta = -8.9\,e9\,m^5/F/coul^2$, $\gamma = +4.5\,e11\,m^9/F/coul^4$ [15]. The dashed line shows the negative dP/dE region which is normally unstable, but is effectively stabilized when placed is series with a normal capacitor.

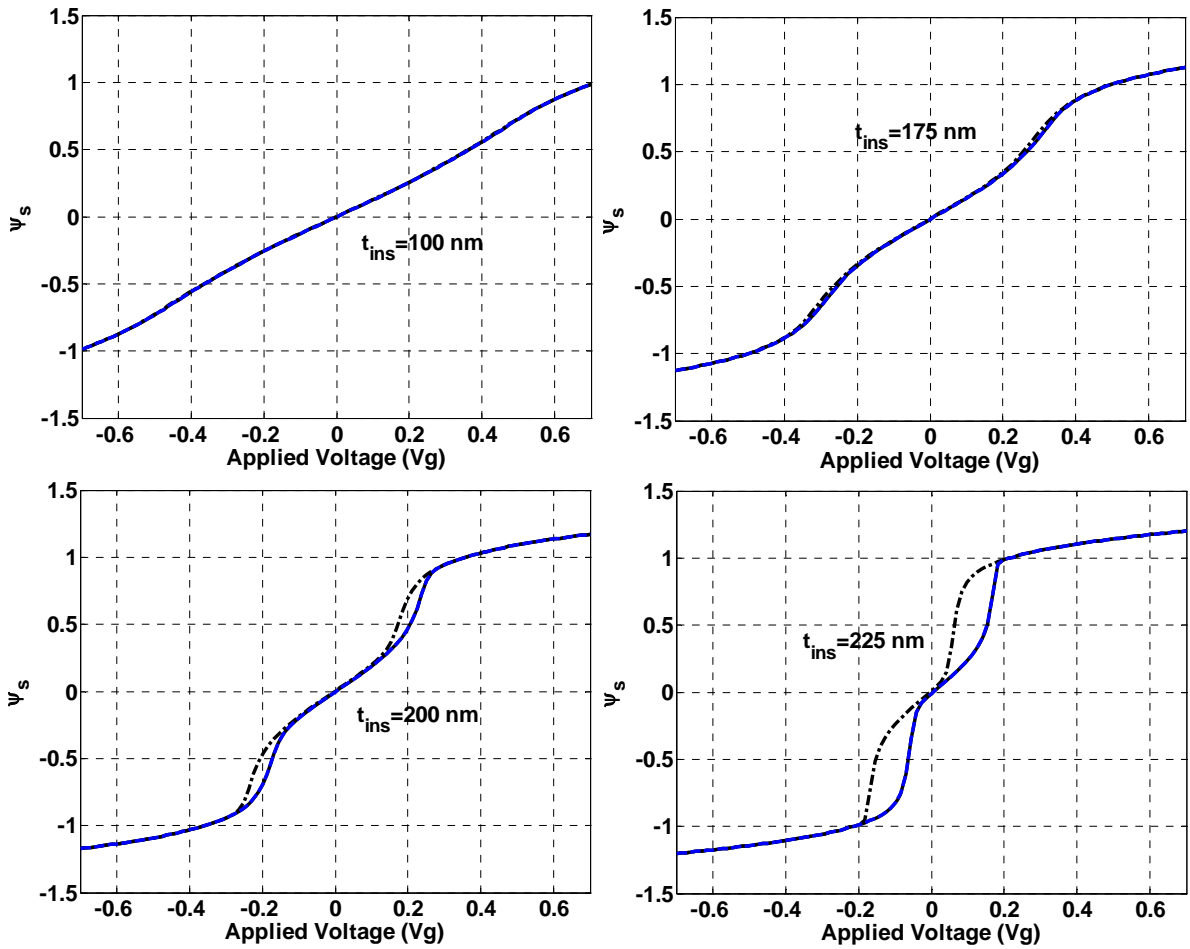

**Fig.3**. The $\psi_s$ vs. $V_g$ plots for different insulator thicknesses. The blue (solid) lines show the sweep from negative to positive voltage and the black (dashed) curve shows the sweep from positive to negative. As the insulator thickness is increased the $\psi_s$ vs. $V_g$ becomes increasingly steeper and at some point the it starts to open up a hysteresis. For an even larger thickness (not shown) the $\psi_s$ vs. $V_g$ plots will no longer pass through origin and resemble a conventional ferroelectric hysteresis curve. Note the similarity of these theoretical plots for $BaTiO_3$ in series with a normal capacitor $C_s$ with the experimental plots in Ref.17 (Figure 1) for $BaTiO_3$ raised to temperatures above its $T_c \sim 107$ C.

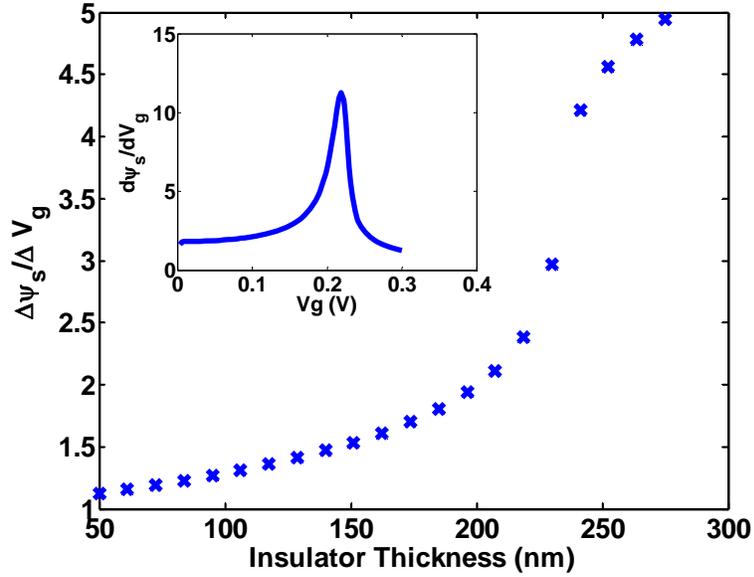

**Fig.4**. $\Delta\Psi_s/\Delta V_g$ calculated from $\Psi_s/V_g$ at $V_g = 0.2$ V, as a function of insulator thickness. In this manner, the $\Delta\Psi_s/\Delta V_g$ reflects the total change of $\Psi_s$ for an applied $V_g = 0.2$ V. The hysteresis starts to manifest itself around 200 nm. It is evident that with a little bit of hysteresis a huge gain in $\Delta\Psi_s/\Delta V_g$ is possible. Note that the actual $d\Psi_s/dV_g$ for specific voltages is many times more than $\Delta\Psi_s/\Delta V_g$ in this voltage range. Inset shows $d\Psi_s/dV_g$ for $t_{ins} = 225$ **nm**.

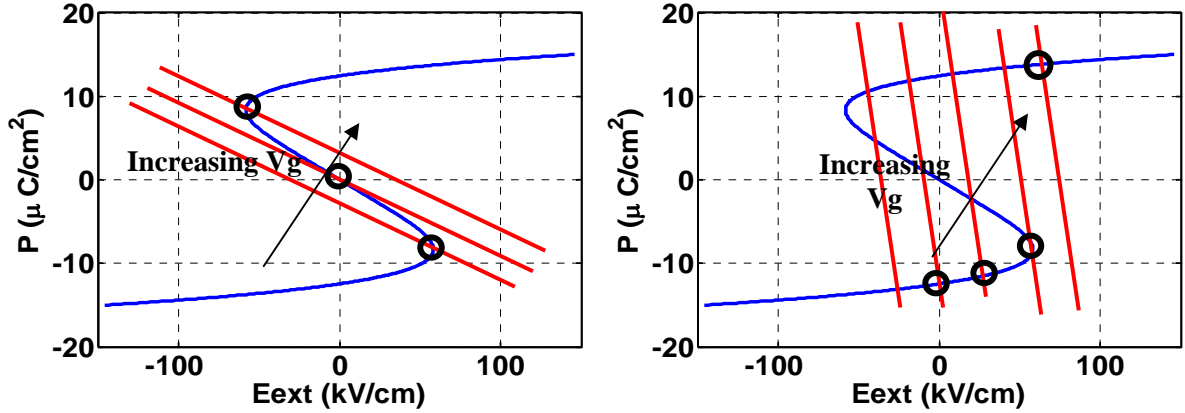

**Fig. S1:** The steady state solution of LK equation gives a S curve. The series capacitance acts as a load line on it. The circles show the operating points when Vg is increasing. A higher slope load line gives a steeper $\partial \psi_s / \partial \mathbf{V_g}$. If the slope of the load line is low, it cuts through the negative slope of the S curve, sampling the curve frequently so that we have $\psi_s$ vs. $V_g$ plot that goes through the origin (see the left panel). When the slope of the load line is large, the $\psi_s$ vs. does not go through origin and a big hysteresis shows up.

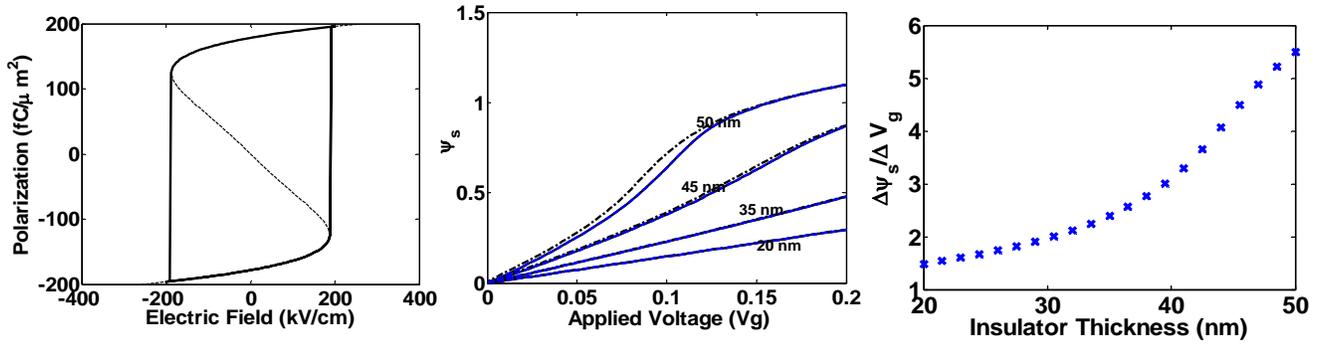

**Fig. S2**. *left panel:* P-E curve for the isolated ferroelectric capacitor. The switching field for the ferroelectric capacitor increases due to the increase in $\alpha$. *middle panel:* The $\psi_s$ vs. $V_g$ plots as a function of insulator thickness. *right panel*: $\Delta\Psi_s/\Delta\mathbf{V_g}$ as a function of insulator thickness. Increasing $\alpha$ circumvents the nonlinear terms and the ferroelectric oxide approaches closer to an ideal negative capacitance. Hence a huge gain in $\Delta\Psi_s/\Delta\mathbf{V_g}$ is possible with almost negligible hysteresis.

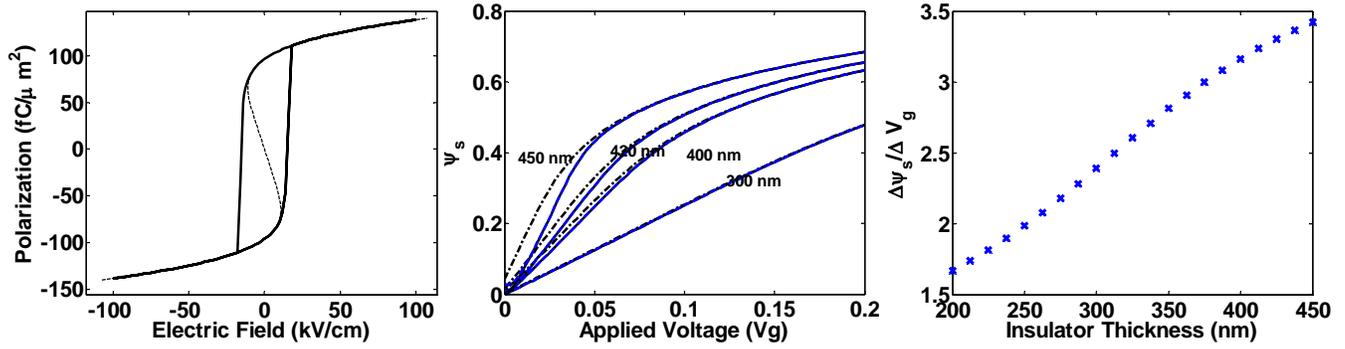

**Fig. S3**. *left panel:* P-E curve for the isolated ferroelectric capacitor. The switching field for the ferroelectric capacitor decreases due to decrease in $\beta$. *middle panel:* The $\psi_s$ vs. $V_g$ plots as a function of insulator thickness. *right panel*: $\Delta\Psi_s/\Delta V_g$ as a function of insulator thickness. the gain in $\Delta\Psi_s/\Delta V_g$ is not that much even when the hsyteresis is such that the $\psi_s$ vs. $V_g$ plot does not go through zero.